\documentclass[aps,prd,preprint,groupedaddress,nofootinbib]{revtex4-1}

\usepackage{graphicx}
\usepackage[table,xcdraw]{xcolor}
\usepackage{xcolor}
\usepackage{amsmath}
\begin{document}

\title{\textbf{Inhomogeneity of a rotating quark-gluon plasma from holography}}

\author{Nelson R. F. Braga\thanks{\href{mailto:nrfbraga@gmail.com}{nrfbraga@gmail.com}},\, 
Octavio.~C. Junqueira\thanks{\href{mailto:octaviojunqueira@gmail.com}{octaviojunqueira@gmail.com}} }
\affiliation{Universidade Federal do Rio de Janeiro---Instituto de F\'isica,  CEP: 21941-909, Rio de Janeiro, RJ,  Brazil}

%\date{}
%\maketitle

\begin{abstract}
 
Rotation affects the transition temperature between confined (hadronic) and deconfined (quark-gluon plasma) phases of the strongly interacting matter produced in non-central heavy ion collisions.  A holographic description of this effect was presented recently, considering an AdS black hole with cylindrical symmetry in rotation. Here we extend this approach in order to analyse the more realistic case of strongly interacting matter that, rather than living in a cylindrical shell, spreads over a region around the rotational axis. In this case, the confined and deconfined phases may coexist.  The holographic description of the plasma behaviour under rotation is shown to be consistent with the concept of local temperature for rotating frames developed by Tolman and Ehrenfest.

\end{abstract}

\maketitle

%\title{\textbf{Spontaneous symmetry breaking in the Donaldson-Witten topological gauge theory }}

\section{Introduction}
The Quark-Gluon Plasma (QGP) is a state of matter formed by deconfined partons that interact strongly \cite{Busza:2018rrf}. Such a state can be created in relativistic heavy-ion collisions produced in particle accelerators.  When the collision is non-central, the resulting quantum chromodynamics (QCD) matter acquires angular momentum. This effect can be indirectly measured since the angular momentum affects the polarization of strange quarks, see \cite{Liang:2004ph}, and induce measurable spin polarization of hadrons \cite{Huang:2020dtn}. Experimental results of RHIC collaboration indicates that this angular momentum is of the order of $10^3\hbar$, with angular velocity $\omega \sim 0.03 \, \text{fm}^{-1}c$ \cite{STAR:2017ckg}. Recent models predict that it could reach even higher values, with $\omega \sim 0.1\,\text{fm}^{-1}c$ \cite{Deng:2016gyh, Jiang:2016woz}. 

Rotation affects the properties of QCD matter, including the critical temperature of the confinement/deconfinement transition between hadronic and plasma phases. For previous studies about the rotation effects in the QGP, see for example \cite{Miranda:2014vaa,Jiang:2016wvv,McInnes:2016dwk,Mamani:2018qzl,Wang:2018sur,Chernodub:2020qah,Arefeva:2020jvo,Chen:2020ath,Zhou:2021sdy,Braguta:2021jgn,Braguta:2021ucr,Golubtsova:2021agl,Fujimoto:2021xix,Braga:2022yfe,Chen:2022mhf,Golubtsova:2022ldm,Zhao:2022uxc,Chernodub:2022}. 
The rotating QCD matter can be investigated using holographic models \cite{Zhou:2021sdy,Braga:2022yfe}, with a rotating AdS black hole (BH) being the holographic dual of the plasma. Such models predict that  rotation influences the critical temperature of phase transition: it decreases with increasing angular velocity, in agreement with recent studies of rotating QGP using phenomenological models \cite{Chen:2020ath,Chernodub:2020qah,Fujimoto:2021xix}. The same behaviour was found in the Nambu-Jona-Lansinio model \cite{Jiang:2016wvv}, in which  the critical temperature decreases due to suppression of the chiral condensate. In contrast, lattice simulations for the phase transition in gluondynamics with relativistic rotation \cite{Braguta:2021jgn,Braguta:2021ucr} indicates that the critical temperature increases with increasing angular velocity.

The inhomogeneity of rotating QCD matter was studied recently in \cite{Chernodub:2020qah} considering a toy model based on the compact Quantum Electrodynamics (QED) theory. 
An outcome of this work is that  that the plasma can present confined and deconfined phases, spatially separated  according to the rotational velocity.  

In the present work we consider the holographic description of a plasma, with arbitrary shape, in a rigid rotation around a fixed axis. The cylindrical slices of the plasma, consisting of the points  that share the same distance to the axis, have the same velocity. Each of these slices can be represented by a rotating anti-de Sitter black hole (AdS BH) with cylindrical symmetry, as it was done in Ref. \cite{Braga:2022yfe}.
Considering a  soft-wall AdS/QCD background we analyse the confinement/deconfinement transition of the rotating plasma. It is found that, depending on the temperature, the angular velocity and the maximum radius of the plasma, the hadronic and plasma phases may coexist, characterizing an inhomogeneity of the medium.

The paper is organized as follows: in Section II, we present the rotating AdS BH geometry and discuss its properties, like temperature.   In Section III, we discuss the confinement/deconfinement phase transition in the holographic soft-wall model. Section IV is devoted to describe the inhomogeneity of QCD matter as an effect of  rotation. Section V contains the comparison between the approach based on the Tolman-Eherenfest law and the holographic description, and, finally, Sections VI contains our conclusions.

\section{Rotating AdS black hole with cylindrical symmetry}
The metric for an AdS black hole with cylindrical symmetry and radius $\rho$, rotating with an angular velocity $\omega$,  is given by \cite{Zhou:2021sdy}

\begin{eqnarray}\label{RotatingBHmetric}
	ds^2 &= & g_{tt} dt^2 + g_{t\phi}dt d\phi + g_{\phi t} d\phi dt + g_{\phi \phi} d\phi^2 \cr && + \,  g_{zz} dz^2 + g_{xx} \sum_{i=1}^2 dx_i^2\;,
\end{eqnarray}
with
\begin{eqnarray}
	g_{tt} &=& \frac{\gamma^2(\omega)L^2}{z^2}\left( \omega^2 \rho^2-f(z) \right)\;, \label{RotatingBHi} \\
	g_{\phi \phi} &=& \frac{\gamma^2(\omega)L^2}{z^2} \rho^2 \left( 1-\omega^2 \rho^2f(z) \right)\;,\\
	g_{t \phi} &=& g_{\phi t} = \frac{\gamma^2(\omega)L^2}{z^2}\omega \rho^2 \left( 1-f(z) \right)\;,\\
	g_{zz} &=& \frac{L^2}{z^2 f(z)}\;,\\
	g_{xx} &=& \frac{L^2}{z^2}\;,\label{RotatingBHf}
\end{eqnarray}
where $f(z) = 1 - z^4/z_h^4$, being $z_h$ the black hole horizon position, $\gamma$ is the Lorentz factor,
\begin{equation}
	\gamma(\omega \rho ) = \frac{1}{\sqrt{1-\omega^2\rho^2}}\;,
\end{equation}
and $\phi$ is the angular coordinate with periodicity $0 < \phi < 2\pi/\gamma$. 

From gauge/gravity duality, this AdS black hole represents a rotating QGP with the form of a cylindrical surface of radius $\rho $.  This simple situation, where the speed is the same for all points of the plasma, is an important tool for representing a realistic plasma, by just considering a superposition of BHs with different radius $\rho $.  
 
Other solution of Einstein's equation in the case with rotation is the thermal AdS space, that has the same form of equations \eqref{RotatingBHi}-\eqref{RotatingBHf} but taking $f(z) =1$. Such a space is used to eliminate ultraviolet divergencies in the Hawking-Page method to determine the critical temperature of confinement/deconfinement transition, as we shall see in the next section.  

After a straightforward calculation, one can rewrite the rotating BH metric \eqref{RotatingBHi}-\eqref{RotatingBHf} in the canonical form:
\begin{eqnarray}\label{canon}
	ds^2 &=& -N(z) dt^2 + \frac{L^2}{z^2}\frac{dz^2}{f(z)} + R(z)\left( d\phi + P(z) dt\right)^2    + \frac{L^2}{z^2} \sum_{i-1}^2 dx_i^2\;,
\end{eqnarray}
with 
\begin{eqnarray}
	N(z) &=& \frac{L^2}{z^2} \frac{ f(z) (1-\omega^2 \rho^2)}{1- f(z)\omega^2 \rho^2}\;, \label{N}\\
	R(z) &=& \frac{L^2}{z^2}\left( \gamma^2 \rho^2 -  f(z) \gamma^2 \omega^2 \rho^4\right)\;, \label{R}\\
	P(z) &=& \frac{\omega(1-f(z))}{1- f(z)\omega^2 \rho^2}\;\label{P}.
\end{eqnarray}
Thus, defining $h_{00} = - N(z)$, the  Hawking temperature can be obtained from the surface gravity formula \cite{Zhou:2021sdy}:
\begin{eqnarray}\label{HT}
	T &=& \vert \frac{\kappa_G}{2\pi} \vert = \bigg\vert \frac{\lim_{z\rightarrow z_h}- \frac{1}{2} \sqrt{\frac{g^{zz}}{-h_{00}(z)}}h_{00,z}}{2\pi}\bigg \vert = \frac{1}{\pi z_h} \sqrt{1-\omega^2 \rho^2}\;, 
\end{eqnarray}
where $\kappa_G$ is the surface gravity, and $g^{zz}$ the $zz$ component of the inverse of the rotating BH metric. Note that if one  takes $\omega \rho \rightarrow 0$, one obtains the temperature 
$ T(\omega \rho \rightarrow 0) = 1/(\pi z_h)$, which is the well-known Hawking temperature of a non-rotating AdS black hole. In this limit, the rotating BH metric \eqref{RotatingBHmetric}-\eqref{RotatingBHf} is reduced to 
\begin{equation}\label{BHAdS}
ds^2= \frac{L^2}{z^2}\left( f(z) dt^2 + d\overrightarrow{x}^2 +  \rho^2 d\phi^2 + \frac{dz^2}{f(z)} \right)\;,
\end{equation}
which is the (non-rotating) cylindrical AdS BH space.  

One can associate this temperature $ 1/(\pi z_h)$ to the local temperature, observed in a frame that is rotating with the plasma and thus observes no rotation. Defining $T_{\text{local}} = 1/(\pi z_h)$ and using the expression for the Hawking temperature \eqref{HT}, one obtains
\begin{equation} \label{Tlocal}
T_{\text{local}} = \frac{T}{\sqrt{1-\omega^2 \rho^2}}\;,
\end{equation}
where $T$ corresponds to the temperature as seen  by a rest frame  that observes the plasma rotating with a rotational velocity $\omega \rho$. 

\section{Hawking-Page transition with rotation in the soft-wall AdS/QCD model}

In the soft-wall holographic AdS/QCD model \cite{Karch:2006pv}, one introduces an energy
parameter in the AdS geometry, which works as an infrared (IR) cutoff in the gauge theory side of the gauge/gravity duality. The five-dimensional gravitational action for this model, at zero temperature, takes the form \cite{Herzog:2006ra,BallonBayona:2007vp}
\begin{equation}\label{action1}
	I = - \frac{1}{ 2 \kappa^2} \int_0^{\infty} dz\int d^4x \sqrt{g} e^{-\Phi}\left( R - \Lambda \right) \; =  \frac{4}{L^2\kappa^2} \int_0^{\infty} dz \int d^4x \sqrt{g} e^{-cz^2},     
\end{equation}
where $\kappa$ is the gravitational coupling associated with the Newton constant, and $\Lambda$, the cosmological constant that is related to the curvature and the AdS radius by  $\Lambda = \frac{3}{5} R = -\frac{12}{L^2}$. The scalar field $\Phi(z)$ is the dilaton background that introduces the IR energy parameter $\sqrt{c}$ in the model. 

Considering the metric \eqref{RotatingBHmetric} for the rotating BH case and  a similar expression,  but with $ f(z) = 1 $, for the thermal AdS one,  the determinant of the metric in both cases is $g = \frac{L^{10}}{z^{10}}$, such that the finite temperature version of \eqref{action1} reads
\begin{equation}
	I_{on-shell} = \frac{4L^3}{\kappa^2} \int_0^{z_f} dz\int d^4x z^{-5}e^{-cz^2}\;\,,
\end{equation} 
where $z_f = z_h$ for the black hole. For the thermal AdS space there is no horizon, so $z_f \rightarrow \infty$. 

As the integration over the spatial bulk coordinates is trivial, we define an action density, $\mathcal{E} = \frac{1}{V_{3D}} I_{on-shell}$, where $V_{3D}$ is the spatial volume factor of the bulk. For a compact time direction in the Euclidean signature with period $\beta_s$ that depends on the space considered, we have $ 0 \leq t < \beta_s $, such that, including the ultraviolet regulator $\varepsilon$, one obtains the action density
\begin{equation}\label{Es}
	\mathcal{E}_s(\varepsilon) = \frac{4L^3}{\kappa^2\gamma} \int_0^{\beta_s} dt \int_\varepsilon^{z_f}dz\, z^{-5}e^{-cz^2}  \;,  
\end{equation}
where, for the AdS with a black hole, one has: $\beta_{BH} = \beta = 1/T$ . By requiring that the asymptotic limits of the BH and thermal AdS geometries are the same at $z = \varepsilon$ with $\varepsilon \rightarrow 0$, one finds that for the thermal AdS space the temporal period $\beta_s$ is $\beta_{AdS} = \beta^\prime = \beta \sqrt{f(\varepsilon)}$. Note that the factor $1/\gamma$ appear in $\mathcal{E}_s$ because we are dividing by the volume of the cylindrical surface, for which the angular coordinate has the range $0 < \phi < 2\pi$, whereas for the rotating BH metric one has:  $0 < \phi < 2\pi/\gamma$. 
 
For both geometries the energy densities are infinite in the
limit $\varepsilon \rightarrow 0$. For this reason one defines a regularized action density for the rotating black hole as the difference between the energy densities of the two geometries, namely, 
\begin{equation}\label{DeltaE}
	\bigtriangleup \mathcal{E}(\varepsilon) = \lim_{\varepsilon \rightarrow 0} \left[\mathcal{E}_{BH}(\varepsilon) - \mathcal{E}_{AdS}(\varepsilon) \right]\;, 
\end{equation}
where
\begin{eqnarray}
	\mathcal{E}_{BH}(\varepsilon) &=& \frac{4L^3}{\kappa^2\gamma} \beta \int_\varepsilon^{z_h}dz\, z^{-5}e^{-cz^2}  \;, \label{DeltaEBH}\\
	\mathcal{E}_{AdS}(\varepsilon) &=& \frac{4L^3}{\kappa^2 \gamma} \beta^\prime \int_\varepsilon^{\infty}dz\, z^{-5}e^{-cz^2}  \;. \label{DeltaEAdS}
\end{eqnarray}

This regularized density is finite in the ultraviolet limit and works as a measure of stability of the rotating black hole. When $\bigtriangleup \mathcal{E}$ is positive, the BH is unstable since the action density of the thermal AdS space is smaller that the black hole one. In contrast, if it is negative, the BH is stable. From gauge/gravity duality, such analysis of stability is related to the transition between the confined hadronic phase and the deconfined plasma one, with the black hole corresponding to the QGP plasma
and the thermal AdS to the hadronic phase.  

Replacing \eqref{DeltaEBH} and \eqref{DeltaEAdS} into \eqref{DeltaE}, one obtains
\begin{equation}
	\bigtriangleup \mathcal{E}(z_h ) \, = \, \frac{ L^3 \pi }{\kappa^2  z_h^3 } \bigg[  e^{-c z_h^2} ( - 1 + c z_h^2 ) + \frac{1}{2} + 
	c^2 z_h^4 \text{Ei}\left( -c z_h^2 \right) \bigg] \,.\label{DeltaEBHAdS}
\end{equation}
The critical temperature for the confinement/deconfinement phase transition is obtained from the condition $\bigtriangleup \mathcal{E}(\omega \rho, T_c) = 0$. The dependence of $\bigtriangleup \mathcal{E}$ on the temperature and on the rotational velocity $ \, \omega \rho \,$ is obtained using the Hawking temperature expression \eqref{HT}. In Figure 1, we plot $\bigtriangleup \mathcal{E}$ as a function of $T$, at six fixed values of $\omega \rho$. 
\begin{figure}[!htb]
	\centering
	\includegraphics[scale=0.6]{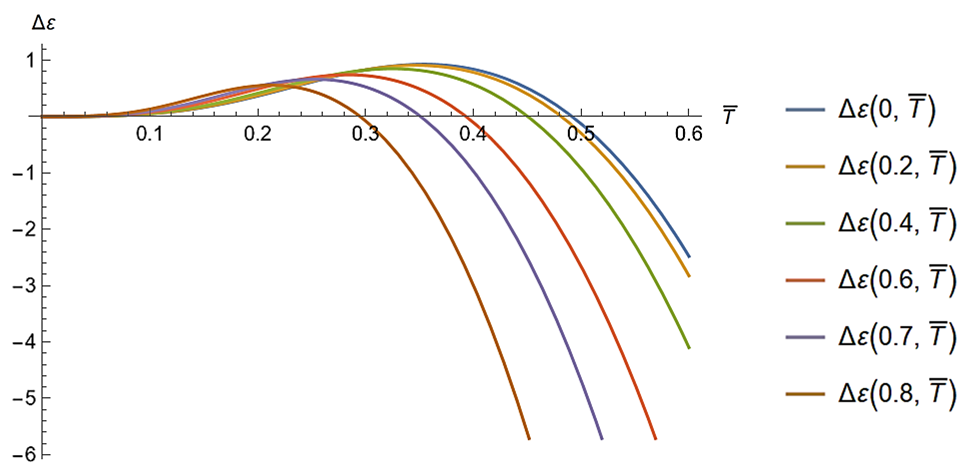}
	\caption{Action density of rotating plasma ($\bigtriangleup \mathcal{E}$) as a function of $\bar{T} = T/\sqrt{c}$, at different rotational velocities $\omega \rho $, and with normalization $L^3/\kappa^2 = 1$. }
\end{figure}

From Eqs. \eqref{HT} and \eqref{DeltaEBHAdS}, one obtains the critical temperature
\begin{equation}
\label{Tcsoft}
   T_c (\omega \rho )  = \frac{ 0.491728 \sqrt{c}}{  \gamma (\omega \rho )}  =  T_c (0) \sqrt{1 -  \omega^2 \rho^2} \,.
\end{equation}
Equation \eqref{Tcsoft} establishes how the critical temperature $T_c$, as observed by a rest frame, depends on the angular velocity and on the rotational radius.
It is interesting to note that the above equation implies that for the rotating plasma the transition occurs at a temperature $ T = T_c (\omega \rho ) =  T_c (0) \sqrt{1 -  \omega^2 \rho^2} $, or equivalently, using Eq. \eqref{Tlocal}, when
\begin{equation}
\label{Tcsoftalternativelocal}
    \frac{ T }{ \sqrt{1 -  \omega^2 \rho^2} } = T_{\text{local}} =  T_c (0) \,,
\end{equation}
or, in other words, when the local temperature is equal to the critical temperature of (non-rotating) QCD matter.

\section{QGP inhomogeneity from holography}

In order to represent the situation that appears in non-central heavy-ion collisions, we consider a region of space with QCD matter rotating with angular velicity $\omega $ around a fixed axis. Using the same notation as in the last sections, we represent the distance to the axis as $\rho$. For a continuous distribution one has $0 < \rho <R$, where $R$ is the maximum  distance to the axis where there is QCD matter. Points located on the axis, for which  $\rho = 0 $, are in the non-rotating rest frame. 

Each portion of QCD matter located at a distance $\rho $ from the axis is represented holographically by a rotating black hole, as the one described in the previous sections, with rotational velocity $ \omega \rho $. From Eq.  \eqref{Tlocal} one notices that the local temperature of this matter distribution increases with $\rho$, reaching a maximum at $\rho = R$,
\begin{equation}
T_{\text{local}}(R) = \frac{T}{\sqrt{1-\omega^2 R^2}}\;,
\end{equation}
such that the local temperature inside the region varies in the range
$T \le T_{local} \le \frac{T}{\sqrt{1-\omega^2 R^2}}$. This result was obtained before in \cite{Chernodub:2020qah} from a different method. 

As expressed in Eq. \eqref{Tcsoftalternativelocal}, the phase transition occurs when $ T_{\text{local}} =  T_c (0)$.  From the previous results, one identifies three possible situations for rotating QCD matter.

\noindent i)  If $T > T_c (0) $, there is just one phase. The QCD matter will be composed of only QGP. 

\noindent ii) If $T_c (0) > T_{\text{local}} (R)$, again there is just one phase.
In this case, the QCD matter will be composed only of hadrons.

\noindent iii)  If $T < T_c (0) < T_{\text{local}} (R)$, the phase transition will occur at a value $\rho_c$ of the radius such that $ T_{\text{local}} (\rho_c)  = T_c (0)  $. In this case
there are two phases: hadronic matter for  $0 < \rho < \rho_c$ and QGP for  $\rho_c <\rho<R$.

  These three configurations are represented in Figure 2.  

\begin{figure}[!htb]
	\centering
	\includegraphics[scale=0.55]{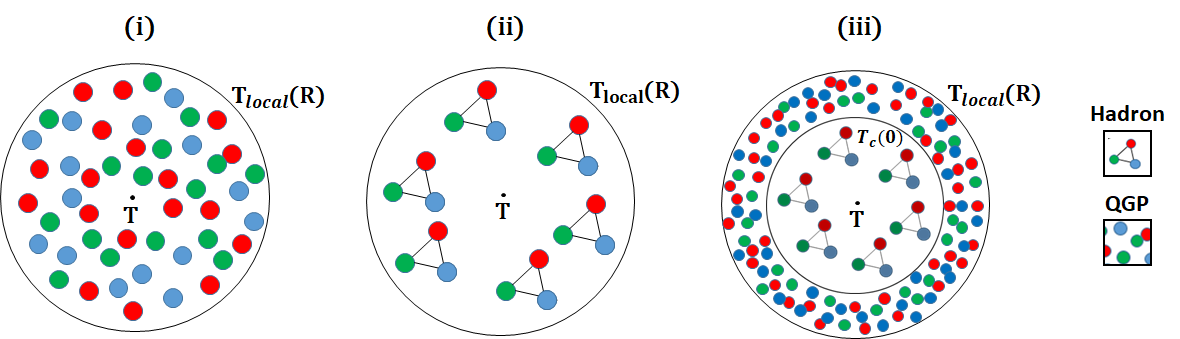}
	\caption{Configurations of rotating QCD matter  in the region $0 < \rho <R$. i) QGP without phase transition for $T > T_c (0)  $. ii) Only hadronic matter  for $T_c (0)  > T_{local} (R)  $. iii)  Phase transition at $\rho_c$ when $ T_{\text{local}} (\rho_c)  = T_c (0)  $.  }
\end{figure}

The description of the existence of such a mixed phase in the plasma in a rigid rotation was 
presented before in \cite{Chernodub:2020qah} using a different approach to determine the critical temperature. The relation between the critical temperature in the rotating and non-rotating frames are given by the same equation in both cases, according to Eq. \eqref{Tcsoft}. Essentially, Figure 2 is equivalent to Figure 7 of reference \cite{Chernodub:2020qah}. 

Such a result shows how the holographic description describes the inhomogeneity of the QCD matter, this phenomenon being driven by the effect of rotation. The configuration (iii) with a phase transition is characterized by the curve of Figure 3, where we have assumed $\omega  = 0.1 \,\text{fm}^{-1}$. If there is no rotation, the inhomogeneity is not produced. The configuration of QCD matter in the mixed phase described by holography possesses a typical signature: the hadrons dominate the inner region near the rotation axis.  

\begin{figure}[!htb]
	\centering
	\includegraphics[scale=0.62]{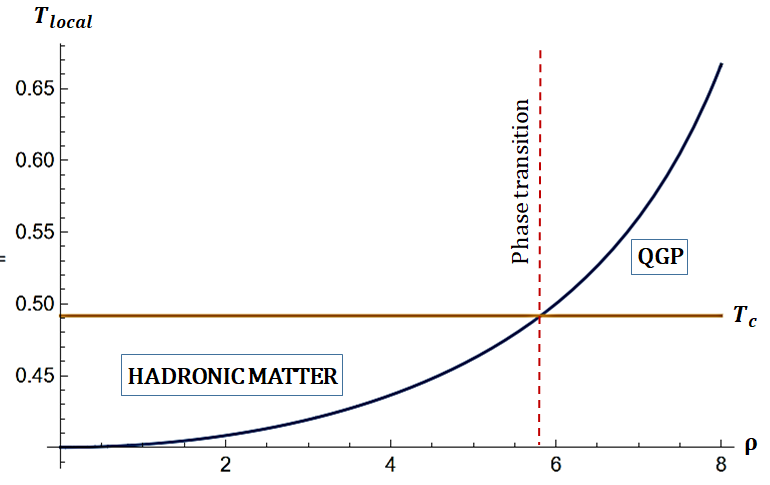}
	\caption{Local temperature of rotating QCD matter versus its radius ($\rho$), assuming $\omega = 0.1 \, \text{fm}^{-1}$, $T = 0.4$ and $T_c = 0.49$, with the energy scale $\sqrt{c} = 1$. The radius is varying from zero to $R = 8 \, \text{fm}$.    }
\end{figure}

\section{Comparison with the Tolman-Ehrenfest effect}

It was argued by  Tolman and Ehrenfest   \cite{Tolman:1930zza, Tolman:1930ona} that for a medium in thermal equilibrium located in a general non-flat space, local observers would measure a temperature that depends on the spacetime metric. In particular, for the rotating metric of eq. \eqref{RotatingBHmetric}  the local temperature of Tolman and Ehrenfest takes the form \cite{Chernodub:2020qah}
\begin{eqnarray}\label{TTE}
    T^{TE}(\rho) = \frac{T^{TE}_0}{\sqrt{1 - \omega^2\rho^2}}\;,
\end{eqnarray}
where  $T^{TE}_0 $ is a reference temperature, that in our case is simply $T$, the temperature as observed by a non-rotating frame, or equivalently the temperature on the rotational axis. Comparing the above result with Eq. \eqref{Tlocal}, one realizes that the local temperature obtained by the holographic description is equivalent to the Tolman-Ehrenfest one.

\section{CONCLUSIONS}

We described the possible inhomogeneity of a region of rotating QCD matter using the holographic soft-wall AdS/QCD model. The rotational radius varies in the range $0 < \rho < R$. For each portion of matter at a given radius $\rho $ there is a particular gravitational dual.  The relation between temperature, rotational velocity and horizon position was determined by the Hawking temperature condition of eq. \eqref{HT}. From this result we found and expression for the local temperature observed by a frame that is at rest with respect to the rotating strongly interacting matter.  As a consequence one can have an inhomogeneous medium, when the temperature at the rotational axis is smaller than the critical temperature in the non-rotating case $T_c (0) $ but the temperature at $ \rho = R $ is greater than $T_c (0) $. It was also shown that the local temperature obtained from holography is equivalent to the Tolman-Ehrenfest one.

The main results obtained here coincide with the ones obtained previously in 
ref. \cite{Chernodub:2020qah} using compact electrodynamics in two spatial dimensions. It is interesting to see that the holographic approach leads to the same behaviour for the rotating plasma. 
\\

\noindent \textbf{Acknowledgments}: The authors are supported by FAPERJ --- Fundação Carlos Chagas Filho de Amparo à Pesquisa do Estado do Rio de Janeiro and CNPq - Conselho Nacional de Desenvolvimento Cient\'ifico e Tecnol\'ogico. This work received also support from  Coordena\c c\~ao de Aperfei\c coamento de Pessoal de N\'ivel Superior - Brasil (CAPES) - Finance Code 001. 

\bibliographystyle{ieeetr}
\bibliography{library}

\end{document}